\documentclass{agujournal2019}
\usepackage{makecell}
\draftfalse

\journalname{JGR: Planets}

\begin{document}

%%%%%%%%%%%%%%%%%%%%%%%%%%%%%%%%%%%%%%%%%%%%%%%
%  TITLE
%
% (A title should be specific, informative, and brief. Use
% abbreviations only if they are defined in the abstract. Titles that
% start with general keywords then specific terms are optimized in
% searches)
%
%%%%%%%%%%%%%%%%%%%%%%%%%%%%%%%%%%%%%%%%%%%%%%%

% Example: \title{This is a test title}

\title{3D Radiative Transfer Simulations for Characterization of the Optically Thick DART Ejecta Plume}

%%%%%%%%%%%%%%%%%%%%%%%%%%%%%%%%%%%%%%%%%%%%%%%
%
%  AUTHORS AND AFFILIATIONS
%
%%%%%%%%%%%%%%%%%%%%%%%%%%%%%%%%%%%%%%%%%%%%%%%

% Authors are individuals who have significantly contributed to the
% research and preparation of the article. Group authors are allowed, if
% each author in the group is separately identified in an appendix.)

% List authors by first name or initial followed by last name and
% separated by commas. Use \affil{} to number affiliations, and
% \thanks{} for author notes.
% Additional author notes should be indicated with \thanks{} (for
% example, for current addresses).

% Example: \authors{A. B. Author\affil{1}\thanks{Current address, Antartica}, B. C. Author\affil{2,3}, and D. E.
% Author\affil{3,4}\thanks{Also funded by Monsanto.}}

\authors{
Yaeji Kim\affil{1}, 
Lioudmila Kolokolova\affil{2}, 
Tony L. Farnham\affil{2}, 
Sabina D. Raducan\affil{3, 4},
Ramin Lolachi\affil{5, 6, 7},
David A. Glenar\affil{5, 6, 7},
Timothy J. Stubbs\affil{7}
}

\affiliation{1}{California State University Long Beach, LongBeach, CA 90840, USA}

\affiliation{2}{University of Maryland, College Park, 4296 Stadium Dr., College Park, MD 20742, USA}

\affiliation{3}{Archaeology, Environmental Changes $\&$ Geochemistry (AMGC), Vrije Universiteit Brussel, Brussels, Belgium}

\affiliation{4}{Reference Systems and Planetology Department, Royal Observatory of Belgium, Brussels, Belgium}

\affiliation{5}{Center for Space Sciences and Technology, University of Maryland, Baltimore County, Baltimore, MD 21250, USA}

\affiliation{6}{Center for Research and Exploration in Space Science and Technology, NASA/GSFC, Greenbelt, MD 20771, USA}

\affiliation{7}{Solar System Exploration Division, NASA Goddard Space Flight Center, Greenbelt, MD, USA}

%(repeat as many times as is necessary)

% Corresponding author mailing address and e-mail address:

% (include name and email addresses of the corresponding author.  More
% than one corresponding author is allowed in this LaTeX file and for
% publication; but only one corresponding author is allowed in our
% editorial system.)

% Example: \correspondingauthor{First and Last Name}{email@address.edu}

\correspondingauthor{Yaeji Kim}{Yaeji.Kim@csulb.edu}

%%%%%%%%%%%%%%%%%%%%%%%%%%%%%%%%%%%%%%%%%%%%%%%
% KEY POINTS
%%%%%%%%%%%%%%%%%%%%%%%%%%%%%%%%%%%%%%%%%%%%%%%
%  List up to three key points (at least one is required)
%  Key Points summarize the main points and conclusions of the article
%  Each must be 140 characters or fewer with no special characters or punctuation and must be complete sentences

% Example:
% \begin{keypoints}
% \item	List up to three key points (at least one is required)
% \item	Key Points summarize the main points and conclusions of the article
% \item	Each must be 140 characters or fewer with no special characters or punctuation and must be complete sentences
% \end{keypoints}

\begin{keypoints}
\item We use 3D Monte Carlo radiative transfer modeling to characterize the optically thick DART plume observed by LICIACube 175 s after impact.
\item A power-law particle size distribution with slope (k=3.0) and a density peak at $\sim$120 m best reproduce near-surface brightness structures.
\item The ejecta mass within 500 m is constrained to $1.25 \times 10^{6}$ kg, consistent with dynamical simulations for the same near-surface region.
\end{keypoints}

%%%%%%%%%%%%%%%%%%%%%%%%%%%%%%%%%%%%%%%%%%%%%%%
%
%  ABSTRACT and PLAIN LANGUAGE SUMMARY
%
% A good Abstract will begin with a short description of the problem
% being addressed, briefly describe the new data or analyses, then
% briefly states the main conclusion(s) and how they are supported and
% uncertainties.

% The Plain Language Summary should be written for a broad audience,
% including journalists and the science-interested public, that will not have 
% a background in your field.
%
% A Plain Language Summary is required in GRL, JGR: Planets, JGR: Biogeosciences,
% JGR: Oceans, G-Cubed, Reviews of Geophysics, and JAMES.
% see http://sharingscience.agu.org/creating-plain-language-summary/)
%
%%%%%%%%%%%%%%%%%%%%%%%%%%%%%%%%%%%%%%%%%%%%%%%

%% \begin{abstract} starts the second page

\begin{abstract}
The Hyperion package is a three-dimensional Monte-Carlo radiative transfer approach to modeling dust ejecta. Because it has proven to be a powerful tool for simulating optically thick environments with complicated structure, we are applying the routines to observations of the DART impact on asteroid Dimorphos. Our analysis focuses on the LICIACube LUKE red-filter image acquired 175 s after impact, when the near-surface plume remained dense and not strongly structured. The model incorporates the updated Dimorphos shape model, spacecraft and solar geometry derived, and size-dependent scattering properties of realistic dust particles. We adopt an axisymmetric conical ejecta geometry and iteratively adjust the particle size distribution (PSD) and vertical number density profile to reproduce observed brightness profiles extracted from the LUKE image. We find that a power-law PSD with slope $k = 3.0$ and a number density peak located at approximately 120 m above the surface best reproduces the brightness contrast between the illuminated surface, the bright ejecta plume, and the adjacent dark band produced by shadowing within the optically thick material. The resulting best-fit model implies a total ejecta mass of $1.25 \times 10^{6}$ kg within 500 m of the impact site. 

\end{abstract}

\section*{Plain Language Summary}
NASA’s DART mission intentionally impacted the small asteroid Dimorphos to test whether a spacecraft could change an asteroid’s orbit. The impact produced a large cloud of dust and debris, called an ejecta plume. Shortly after impact, the Italian LICIACube spacecraft acquired a time-series of plume images. In the earliest images, the dust cloud close to the asteroid was so dense that it blocked and scattered sunlight, creating bright regions and a noticeable dark band near the surface. In this study, we use a three-dimensional radiative transfer model to simulate how sunlight travels through and scatters within this dense dust cloud. By adjusting the sizes and vertical distribution of dust particles in the model, we reproduce the brightness patterns seen in the spacecraft images 175 seconds after impact. Our results suggest that small and large particles play an important role in shaping the observed brightness and shadowing. We estimate that about 1.25 million kilograms of material were present within 500 meters of the asteroid’s surface at that time. This work helps connect what we see in images to the physical properties of the ejecta and improves our understanding of how asteroid impacts redistribute material.

%%%%%%%%%%%%%%%%%%%%%%%%%%%%%%%%%%%%%%%%%%%%%%%
%
%  BODY TEXT
%
%%%%%%%%%%%%%%%%%%%%%%%%%%%%%%%%%%%%%%%%%%%%%%%

%%% Suggested section heads:
% \section{Introduction}
%
% The main text should start with an introduction. Except for short
% manuscripts (such as comments and replies), the text should be divided
% into sections, each with its own heading.

% Headings should be sentence fragments and do not begin with a
% lowercase letter or number. Examples of good headings are:

% \section{Materials and Methods}
% Here is text on Materials and Methods.
%
% \subsection{A descriptive heading about methods}
% More about Methods.
%
% \section{Data} (Or section title might be a descriptive heading about data)
%
% \section{Results} (Or section title might be a descriptive heading about the
% results)
%
% \section{Conclusions}

\section{Introduction} 
\label{Sec:Introduction}
The Double Asteroid Redirection Test (DART) was NASA's first planetary defense mission designed to test the kinetic impactor technique by striking Dimorphos, the secondary of the Didymos binary asteroid system, and measuring the resulting orbital deflection \cite{rivkin2023planetary,daly2023derived}. One of the main mission objectives was to quantify the momentum transfer efficiency of the impact, including the additional recoil imparted by escaping ejecta, commonly expressed through the momentum enhancement factor, $\beta$. Prior to impact, Dimorphos orbited Didymos with a period of approximately 11~h~55~min; following the collision, its orbital period decreased by about 33 minutes, demonstrating a substantial deflection and confirming the importance of ejecta-driven momentum enhancement \cite{thomas2023orbital,naidu2024orbital,scheirich2024dimorphos}.

The properties and evolution of the ejecta plume are therefore directly linked to the impact physics and momentum transfer efficiency. Imaging obtained immediately after impact revealed a highly structured and rapidly evolving ejecta cloud. In addition to the pre-impact DRACO images from the DART spacecraft, the Italian LICIACube CubeSat, contributed by the Italian Space Agency (ASI), performed a close flyby and acquired high-resolution images of the ejecta plume from distinct viewing geometries \cite{dotto2024dimorphos, farnham2025high}. These data show a complex, optically thick plume containing bright ejecta regions and pronounced shadowed structures, including a dark band adjacent to Dimorphos. Interpreting such features requires a quantitative framework that accounts for multiple scattering, attenuation, and three-dimensional geometry.

Radiative transfer provides a framework for describing how radiation propagates through complex media such as dusty environments. As photons pass through these media, they undergo absorption, scattering, and thermal emission processes that are governed by the radiative transfer equation (RTE). Because the RTE is an integro-differential equation without a general analytic solution for realistic three-dimensional geometries, numerical approaches are required \cite{steinacker20033d,steinacker2006ray,dullemond2000efficient,dullemond20022}. In optically thick and spatially structured environments, simplified one-dimensional or optically thin approximations can lead to systematic biases because the radiation field depends sensitively on geometry, multiple scattering, and shadowing effects. A widely adopted numerical approach is the Monte Carlo method, in which the radiation field is represented by discrete photon packets that are tracked as they move through a spatial grid with prescribed density distributions. The interactions of each packet with the medium are determined by random sampling of probability distribution functions for absorption and scattering events. Advances in computational capability now permit fully three-dimensional simulations for complex geometries \cite{wolf1999multidimensional,harries2000synthetic,gordon2001dirty,misselt2001dirty,wood2001scattered,wood2002infrared,wood2002spectral,whitney2003two,whitney2004two,dullemond2004flaring,jonsson2006sunrise,pinte2006monte,min2009radiative}. Open-source tools such as Hyperion \cite{robitaille2011hyperion} provide flexible frameworks for multidimensional radiative transfer modeling.

While three-dimensional radiative transfer has been applied in planetary atmospheres and astrophysical systems, particularly in cases involving strong horizontal inhomogeneity, its use in small-body environments has been more limited. In many cometary or dust-tail applications, the material is assumed to be optically thin and geometrically simple \cite{marschall2020cometary,bockelee2004composition,debout2016radiative}. The DART ejecta plume, however, represents a rare case in which the low-altitude environment is both optically thick and geometrically structured, making it particularly well suited for three-dimensional radiative transfer analysis. In this study, we apply the three-dimensional radiative transfer code Hyperion to model the ejecta plume as captured by the LUKE images acquired 175~s after impact. By comparing synthetic images with the actual observations, we constrain the particle size distribution, vertical number density structure, and total ejecta mass of the near-surface plume. This approach enables a physically consistent interpretation of the observed brightness morphology and provides quantitative insight into the early-time ejecta properties following the DART impact.

This paper is organized as follows. After the Introduction in Section~\ref{Sec:Introduction}, Section~\ref{Sec:Methodology} describes the three-dimensional radiative transfer framework used in this study. Section~\ref{SubSec:Reference LICIACube} introduces the reference LICIACube image employed to constrain the DART ejecta, while Sections~\ref{SubSec:Geometric}–\ref{SubSec:ejecta_structure} detail the adopted geometric configuration, light-scattering properties, and structural assumptions of the ejecta plume, respectively. Section~\ref{Sec:Results} presents the simulation results. Section~\ref{SubSec:PSD} examines the influence of the power-law particle size distribution, Section~\ref{SubSec:Number_density} describes the adjustment of the vertical number density profile, and Section~\ref{SubSec:BrightnessComparison} provides a direct comparison between the synthetic and observed brightness profiles to identify the best-fit solution. Finally, Section~\ref{Sec:Discussion_Conclusions} discusses the implications of these results and summarizes the main conclusions of this work.

\section{Radiative Transfer Model} 
\label{Sec:Methodology}
To analyze the dust plume, we used the radiative transfer code Hyperion to simulate the brightness distribution observed in the LICIACube images. Hyperion is a three-dimensional radiative transfer tool based on the Monte Carlo method, developed by \citeA{robitaille2011hyperion}. It simulates the trajectories of photon packets as they interact with a dusty medium undergoing scattering, absorption, and re-emission, allowing for realistic modeling in geometrically complex environments. A notable strength of Hyperion is its flexibility: it supports irregularly sampled spatial grids, various coordinate systems, and mixtures of particles of variety of sizes and compositions, making it particularly well-suited for representing irregularly shaped asteroids and spatially variable ejecta structures, as required in the DART case. The code also considers the intricate structure of the ejecta cloud, observational and solar illumination geometry, and surface reflections from Dimorphos. This modeling framework has been successfully applied in similar contexts, including the analysis of the ejecta from Comet 9P/Tempel 1, a NASA’s Deep Impact mission target \cite{a2005deep}, as demonstrated in the work by \cite{nagdimunov2014properties}. 

\begin{figure}
\centering
\includegraphics[width=\textwidth]{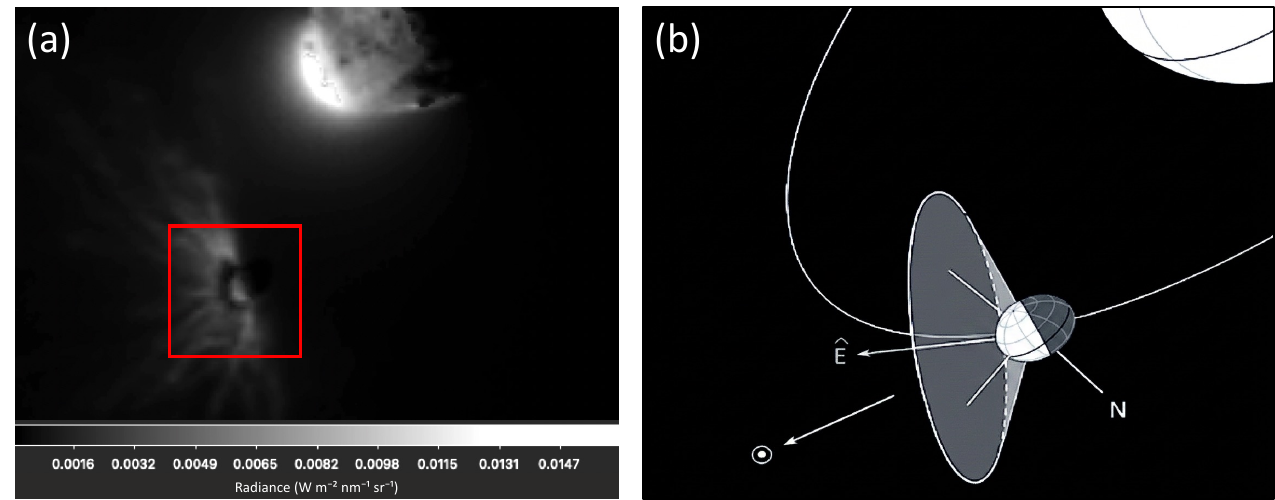}
\caption{(a) LUKE red filter image of the DART ejecta plume acquired 175 s after impact. The rectangular outline shows the region used to generate the Hyperion synthetic image. (b) Schematic representation of the enlarged region from panel (a). The diagram shows the solar illumination direction and the orientation of the ejecta cone relative to  Dimorphos' surface.}
\label{fig:LUKEImage}
\end{figure}

To effectively utilize Hyperion for modeling the DART ejecta, a critical step involves accurately configuring the computational grid to represent the physical state of the system at the specific moment of interest. We focus on the LUKE image taken 175 seconds after the DART impact seen in Figure~\ref{fig:LUKEImage}(a), which captures the major-stage morphology of the ejecta plume. To enable quantitative modeling, we restrict our analysis to projected altitudes within 500~m of the surface (indicated by the rectangular region in Figure~1(a)), where the ejecta remains optically thick and its distribution is comparatively homogeneous. Constructing the model requires precise geometric information, including the relative positions of the Sun, Dimorphos with its expanding ejecta plume, and the LICIACube spacecraft at the time of observation. Additionally, it is essential to define the scattering properties of both the surface of Dimorphos and the ejecta particles. The key free parameters in this modeling effort is the size distribution of the dust particles, assumed to follow a power-law, and the spatial variation of particle number density, which is initially derived from the ejecta mass distribution of \citeA{raducan2024physical}, assuming that the particle composition and scattering properties are known. By iteratively adjusting these parameters to reproduce the brightness profiles observed in the LICIACube images, we place quantitative constraints on the amount and distribution of material ejected by the impact.

\subsection{Reference LICIACube image}
\label{SubSec:Reference LICIACube}
We focus on quantifying the DART ejecta captured by LUKE at a distance of approximately 71 km, 175 s after impact. At this time, the ejecta plume remains optically thick over several hundred meters above the surface of Dimorphos, making it a particularly suitable case for modeling with our 3D radiative transfer approach. For this analysis, we use the calibrated red-channel image provided by \citeA{farnham2025high} (see Figure~\ref{fig:LUKEImage}(a)).\footnote{The calibrated image is archived in the NASA Planetary Data System (PDS) Small Bodies Node; a pre-release version is currently available at \url{https://pdssbn.astro.umd.edu/draft/doi/10.26007/LICIACubeFarhamCalibration.shtml}; see also \citeA{farnham2025luke}.} These data were reduced using improved background removal and radiometric calibration coefficients. The image used in this study is a merged product constructed from three sequential frames with different exposure times. The frames were calibrated to radiance and registered before saturated or near-saturated pixels in the longer exposures were replaced with calibrated, unsaturated pixels from the shorter exposures. This procedure increased the signal-to-noise ratio while minimizing saturation and preserving consistent radiance units. The data reduction, calibration, and image-merging procedures are described in detail by \citeA{farnham2025high}. No additional contrast enhancement or nonlinear intensity transformation was applied to the numerical radiance array used in the present analysis. Figure~\ref{fig:LUKEImage}(a) is displayed using a linear intensity scale, and the brightness profiles were extracted directly from the flux-calibrated merged image in units of W~m$^{-2}$~nm$^{-1}$~sr$^{-1}$. The display settings therefore do not alter the radiance values compared with the Hyperion simulations.

We restrict the modeled region to projected altitudes within 500~m of the Dimorphos surface, corresponding to the rectangular outline in Figure~\ref{fig:LUKEImage}(a). This region contains the near-surface, optically thick plume that is the focus of our analysis. For computational tractability, we approximate its large-scale number density as varying only with height above the surface. The observed ejecta are not actually homogeneous: overlapping filaments, clumps, and other irregular structures are visible in the LUKE image. We therefore seek to reproduce the large-scale radiance structure, particularly the bright ejecta and the prominent dark band near the surface, rather than the complete pixel-scale morphology of the plume.

% We focus on quantifying the DART ejecta captured by LUKE at a distance of approximately 71 km, 175 s after impact. At this time, the ejecta plume remains optically thick over several hundred meters above the surface of Dimorphos, making it a particularly suitable case for modeling with our 3D radiative transfer approach. For this analysis, we use the calibrated image provided by \citeA{farnham2025high} (see Figure~\ref{fig:LUKEImage}(a))\footnote{The calibrated image is archived in the NASA Planetary Data System (PDS) Small Bodies Node; a pre-release version is currently available at \url{https://pdssbn.astro.umd.edu/draft/doi/10.26007/LICIACubeFarhamCalibration.shtml}.}. In the Hyperion modeling, we constrain our domain to projected altitudes within 500 m of the surface corresponding to the rectanglar outline in Figure~\ref{fig:LUKEImage}(a), where the ejecta is optically thick and its distribution relatively homogeneous. This constraint allows us to simplify the modeling by assuming a dust plume with a number density that varies only with height above the surface, thereby enabling us to probe the properties of particles freshly ejected from Dimorphos. Although the ejecta structure is intrinsically complex, with overlapping filaments and irregular features evident in the image, we focus on the brightest ejecta and the prominent dark band near the surface, rather than attempting to reproduce the full morphological complexity of the plume.

\subsection{Geometric parameters setting}
\label{SubSec:Geometric}
To define the geometric positions of Dimorphos relative to the Sun, the Sun's location, and LICIACube relative to Dimorphos, we used the SPICE kernels from the DART metakernel \textit{dart$\_$tlf$\_$v01.tm} (\url{https://naif.jpl.nasa.gov/pub/naif/pds/pds4/dart/dart_spice/spice_kernels/mk/}). Figure~\ref{fig:LUKEImage}(b) provides a schematic overview of the illumination and viewing geometry at approximately 175~s after impact, serving as a reference for the spatial orientation of the ejecta plume in the radiative transfer modeling. The relative positions of the Sun and LICIACube with respect to Dimorphos, as well as the orientation of the ejecta plume, were derived from LUKE images in combination with the updated DART trajectory information now available in the SPICE kernels \cite{farnham2025high}. All geometric quantities were computed in the Dimorphos-fixed reference frame using spherical coordinates, where $\phi$ corresponds to east longitude and $\theta$ is measured from the north pole, such that $\theta = 90^\circ - \mathrm{latitude}$. 

For the LUKE image acquired 175~s after impact, the observer (LICIACube) was located at $\phi_{\mathrm{obs}} = 127.0^\circ$ and $\theta_{\mathrm{obs}} = 122.9^\circ$ (latitude $ =-30.0^\circ$). At the same epoch, the Sun was located at $\phi_{\mathrm{Sun}} = 218.1^\circ$ and $\theta_{\mathrm{Sun}} = 98.5^\circ$ (latitude $= -9.5^\circ$), while the nominal plume axis was oriented toward $\phi_{\mathrm{plume}} = 264.3^\circ$ and $\theta_{\mathrm{plume}} = 98.8^\circ$ (latitude $= -8.8^\circ$). Figure~\ref{fig:LUKEImage}(b) presents a schematic illustration of this configuration, showing the relative positions of the Sun, Dimorphos, and the ejecta plume as observed by LICIACube 175~s after the DART impact. The apex of the conical ejecta plume is located at the DART impact site on the surface of Dimorphos and defines the origin of the plume axis direction in the Dimorphos-fixed frame.

In the Hyperion simulations, solar illumination was implemented using a parallel-plane light source configuration, in which the Sun is treated as a distant source producing parallel rays across the Dimorphos system. The illumination direction is specified by the Sun’s position in the Dimorphos-fixed spherical coordinate system, while the source distance is effectively assumed to be infinite, consistent with the actual Sun–Dimorphos separation. The incident solar flux was computed from the solar luminosity scaled to the heliocentric distance of Dimorphos at the time of observation, using $F_\odot = L_\odot / (4\pi d_{\mathrm{sun}}^2)$, where $d_{\mathrm{sun}} = 1.56 \times 10^8\ \mathrm{km}$ is the Sun--Dimorphos distance. The finite angular diameter of the Sun ($\sim0.5^\circ$) introduces only negligible deviations from perfect parallel illumination at the spatial scales considered here. This approximation accurately represents solar illumination for kilometer-scale modeling of the ejecta plume and surface interactions.

The shape of Dimorphos was represented using version~3.0 of the DART Dimorphos shape model generated with the stereophotoclinometry pipeline (SPC; \cite{barnouin2020digital,palmer2022practical}) This shape model, presented by \citeA{daly2023derived}, was reconstructed using stereophotoclinometry applied to DART terminal-phase imaging. The SPC approach combines stereophotogrammetry and photoclinometry to iteratively refine surface topography, yielding a high-resolution, facet-based model that captures Dimorphos’s global shape and large-scale boulder distributions. This model provides the geometric foundation for computing accurate illumination and viewing geometries, as well as ejecta–surface interactions, in the Hyperion radiative transfer simulations.

\subsection{Light scattering properties setting}
\label{SubSec:Light_scattering}
Hyperion provides two principal approaches for treating dust scattering: (1) an analytic Henyey–Greenstein dust model, in which the phase function is parameterized by a single asymmetry parameter, and (2) a tabulated dust model referred to by \citeA{robitaille2011hyperion} as the ``spherical particle model" in which the full scattering properties are computed for homogeneous spherical grains using Mie theory. In this study, we adopt the spherical particle model to represent the DART ejecta, but we do not assume spherical particles. In the original implementation described by \citeA{robitaille2011hyperion}, the wavelength-dependent absorption and scattering efficiencies, single-scattering albedo, and angular phase functions are derived from Mie theory for homogeneous spheres. Here, we extend this framework by replacing the Mie-computed optical properties with precomputed scattering parameters for irregular particles. Specifically, following \citeA{lolachi2025ejecta}, the light-scattering properties were derived from discrete dipole approximation (DDA) calculations and laboratory measurements of irregular silicate particles from the Granada–Amsterdam light-scattering database (e.g., \citeA{munoz2020experimental, munoz2025granada}). These data provide extinction and scattering efficiencies, single-scattering albedos, and full angular phase functions appropriate for silicate irregular particles. These tabulated properties are supplied directly to Hyperion as inputs, allowing the radiative transfer solution to incorporate realistic particle scattering behavior. For centimeter-scale fragments not represented in the adopted scattering database, we use the \(H,G_1,G_2\) photometric model of \citeA{muinonen2010three}, with parameters constrained by the Didymos phase-function measurements of \citeA{hasselmann2024unusual}. Following \citeA{lolachi2025ejecta}, this treatment is adopted as an empirical large-particle limit for centimeter-scale and larger ejecta. We recognize that disk-integrated \(H,G_1,G_2\) parameters may include macroscopic surface roughness and self-shadowing effects that are not directly equivalent to scattering by isolated fragments. Nevertheless, within the adopted PSD, the optical signal is dominated by smaller particles, whereas the centimeter-scale population contributes comparatively modestly to the scattered radiance. Consequently, this approximation is not expected to materially affect the inferred large-scale plume structure, although its empirical nature should be kept in mind when interpreting the PSD and number-density normalization.

This tabulated-dust approach is more appropriate for the DART plume because the inferred particle size distribution spans sub-micron to centimeter scales, with a significant contribution from large grains. Unlike the Henyey–Greenstein parameterization, which approximates angular scattering with a prescribed phase function, the tabulated approach retains wavelength-dependent extinction and scattering efficiencies, single-scattering albedo, and full phase functions for each grain size. This treatment provides a physically consistent description of radiative transfer within an optically thick, poly-dispersed ejecta plume. Following \citeA{lolachi2025ejecta,lolachi2023optical}, we defined the light-scattering properties for representative particle radii of 0.5~$\mu$m, 5~$\mu$m, 125~$\mu$m, 5~mm, and 5~cm. For each size, the single-scattering albedo, extinction and scattering efficiencies, and angular phase functions were supplied as inputs to Hyperion. The bulk scattering behavior of the ejecta plume is obtained by combining these discrete grain populations according to the adopted particle size distribution.

To account for the size distribution, the scattering properties are weighted using a differential power-law distribution of the form
\[
n(r) \propto r^{-k},
\]
where $n(r)$ is the differential number density of particles with radius $r$, and $k$ is the power-law size distribution index. Using this weighting, we define the effective single-scattering albedo, the mass extinction parameter $\chi$, and the averaged phase functions for brightness and polarization.

The mass extinction parameter $\chi$ is defined as
\begin{equation}
    \chi = \frac{C_{\mathrm{ext}}}{m},
\end{equation}
where $C_{\mathrm{ext}}$ is the extinction cross section of an individual particle and $m$ is its mass. The particle mass is determined from the particle radius $r$ and grain density $\rho$, assuming $\rho = 3500\,\mathrm{kg\,m^{-3}}$ \cite{britt2003stony}.

The size-averaged scattering properties used in the Hyperion simulations are expressed as
\begin{equation}
    \langle \mathrm{SSA} \rangle =
    \frac{\sum_i C_{\mathrm{sca},i}\, r_i^{-k}}
         {\sum_i C_{\mathrm{ext},i}\, r_i^{-k}},
\end{equation}

\begin{equation}
    \langle \chi \rangle =
    \frac{\sum_i C_{\mathrm{ext},i}\, r_i^{-k}}
         {\sum_i \rho \,\frac{4}{3}\pi r_i^{3}\, r_i^{-k}},
\end{equation}

\begin{equation}
    \langle P_1 \rangle =
    \frac{\sum_i P_{1,i}\, r_i^{-k}}
         {\sum_i r_i^{-k}},
\end{equation}

\begin{equation}
    \langle P_2 \rangle =
    \frac{\sum_i P_{2,i}\, r_i^{-k}}
         {\sum_i r_i^{-k}},
\end{equation}

\noindent
where $C_{\mathrm{sca},i}$ and $C_{\mathrm{ext},i}$ are the scattering and extinction cross sections for particles of radius $r_i$, respectively; $P_{1,i}$ and $P_{2,i}$ are the brightness and polarization phase functions; and $\rho$ is the particle bulk density.

For the surface of Dimorphos, we adopted the Henyey-Greenstein (HG) phase function in the Hyperion modeling to represent its surface reflectance. The HG model in Hyperion is well suited for approximating the surface scattering behavior of small bodies with relatively low albedo, such as Dimorphos, and has been applied in previous studies. In particular, the HG formulation was successfully used in earlier Hyperion modeling of the Deep Impact target, Comet 9P/Tempel 1 \cite{nagdimunov2014properties}, ensuring consistency and comparability between studies. 

For Dimorphos, we initially adopted an estimate of the single-scattering albedo of 0.126, as reported by \citeA{buratti2024pre}, which is lower than that typically measured for other S-type asteroids. To reproduce the LICIACube images more accurately, we adjusted this parameter and found that a single-scattering albedo of 0.10 provided the best match to reproduce the observed brightness, although it is slightly lower than the reported estimate.

A similar situation was reported in radiative transfer studies of the Deep Impact ejecta, where a single-scattering albedo of 0.065 provided a better match to the observed brightness than the value of 0.043 previously derived from Hapke modeling \cite{li2007deep,li2013photometric}. Such discrepancies are not unexpected, as classical radiative transfer approaches often perform poorly when applied to densely packed media such as cometary or asteroid surfaces. Accordingly, our result should not be regarded as a precise measurement of the physical albedo of Dimorphos, but rather as an effective albedo within the range reported by other studies that reproduce the photometric properties observed in the LICIACube images under the framework of our radiative transfer modeling. The model includes direct solar illumination of Dimorphos surface facets that are not occulted by the body or the ejecta. Photons scattered from the modeled Dimorphos surface into the ejecta are also included in the radiative-transfer calculation. Didymos itself is not included as a secondary illuminating source. However, \citeA{lolachi2025ejecta} estimated this contribution under the LICIACube observing geometry and found that it produces less than 1\% of the radiance generated by direct solar scattering for particles up to millimeter sizes, which dominate the optical signal in the adopted PSD. We therefore expect indirect illumination from Didymos to have a negligible effect on the modeled surface--shadow--plume contrast.

To enable direct comparison between the radiative transfer outputs and the LICIACube observation, the modeled radiance values (W\,m$^{-2}$\,sr$^{-1}$\,Hz$^{-1}$) were converted into the same units and image geometry as those of the LUKE camera. The LICIACube LUKE instrument records calibrated radiance in physical units of W m$^{-2}$ nm$^{-1}$ sr$^{-1}$, and all model results were therefore converted to these units to maintain consistency with the observational data. The comparison was performed using the red-filter images, which provide the highest signal-to-noise ratio and best capture the ejecta plume morphology. In this work, we approximate the LUKE red filter by its central wavelength \cite{poggiali2022expected}, as uncertainties in the particle size distribution exceed the radiometric variation introduced by the finite filter bandwidth. This radiometric and geometric calibration ensures that the simulated brightness distributions from Hyperion can be quantitatively compared with the LICIACube observation within a consistent spatial and photometric framework.

\subsection{DART ejecta structure setting}
\label{SubSec:ejecta_structure}
We represent the ejecta morphology with a simple, axisymmetric conical plume of full opening angle 115$^{\circ}$ and inner angle 105$^{\circ}$, adopting a density distribution that varies with height above the surface along the cone. This parameterization is intended to capture the large-scale geometry of the plume within $\sim$500 m of the surface, where the ejecta remain optically thick and local inhomogeneities are effectively averaged in the radiative transfer solution.
Published analyses of LICIACube imaging and numerical impact simulations describe the DART ejecta as broadly conical, while also indicating time-dependent evolution, azimuthal variability, and internal structure such as filaments and clumps \cite{dotto2024dimorphos,ferrari2025morphology,raducan2024physical,deshapriya2023characterization,hirabayashi2025elliptical}. Reported opening angles and plume morphologies vary with epoch and modeling approach, and some studies suggest departures from perfect axisymmetry. In this context, the adopted 115$^{\circ}$ axisymmetric cone should be regarded as an intermediate representation of the large-scale plume structure at the epoch of interest. Because our analysis focuses on the near-surface, optically thick region rather than detailed filamentary morphology, a simplified conical geometry provides an adequate and computationally tractable description for the radiative transfer modeling.

\section{Results} 
\label{Sec:Results}
To constrain the ejecta mass and particle properties, we compare brightness profiles extracted from the observed LUKE red-filter image with those derived from Hyperion-simulated images. By iteratively adjusting the particle size distribution (PSD) and the number density profile within the radiative transfer framework, we identify the parameter combinations that best reproduce the observed brightness structure of the near-asteroid ejecta plume.

\subsection{Power-law Size Distribution adjustment}
\label{SubSec:PSD}
We first investigate the influence of the particle size distribution (PSD) on the modeled brightness structure. Previous studies provide important guidance on the plausible PSD of the DART ejecta. Modeling of the post-impact dust tail by \cite{moreno2023characterization} showed that HST and ground-based observations are best matched by a broken power law, with a slope of approximately $-2.5$ for particle sizes $\leq 3$~mm and a steeper slope of $-3.7$ for larger grains extending up to several centimeters. This result indicates that while micron- to millimeter-scale grains dominate the scattering cross section, the ejecta mass is primarily carried by the largest fragments. Similarly, \citeA{li2023ejecta} used HST observations over an 18.5-day interval to derive PSD slopes of $-2.6 \pm 0.2$ for micron-to-millimeter grains and $-3.6 \pm 0.2$ for millimeter-to-centimeter particles, consistent with the progressive evolution of the observed size distribution over time. 

Early-phase constraints from LICIACube imaging provide additional context. \citeA{lolachi2025ejecta} derived PSD properties from radiative transfer modeling of the ejecta brightness, finding that a single power-law PSD with slope $k \approx 2.5$ (where $n(r) \propto r^{-k}$) provides a good fit to the observations, while a broken power-law formulation is not favored. Their analysis requires particle sizes spanning from sub-micron to $\sim$1~cm.

\begin{figure*}[ht!]
\centering
    \includegraphics[width=\textwidth]{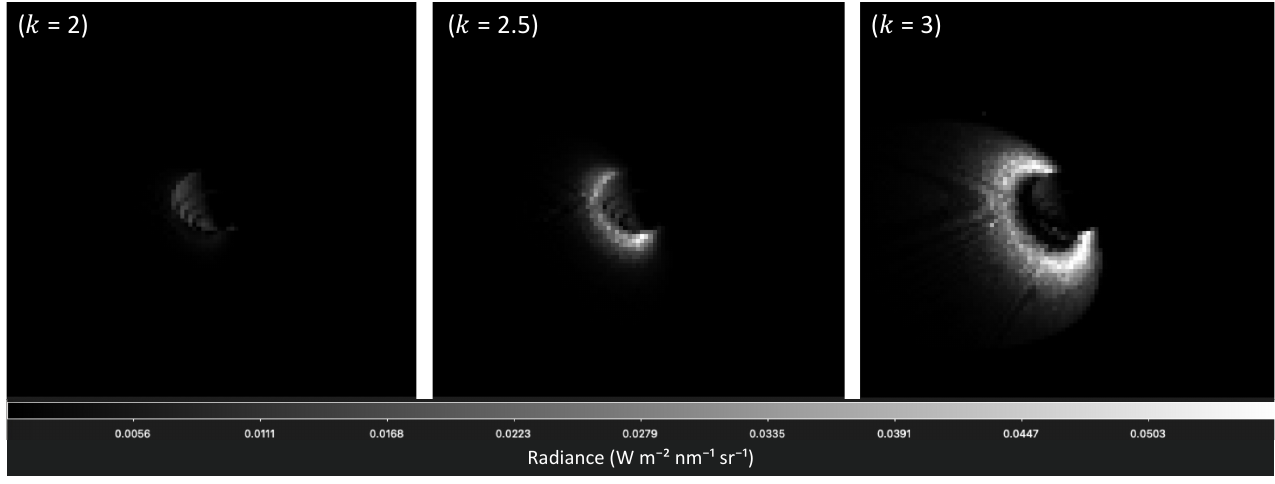}
    \caption{Hyperion-simulated images depending on different size distribution index ($k$)}
    \label{fig:left}
\label{fig:size_distribution}
\end{figure*}

Within this framework, our chosen test slopes of $k = 2.0$, $2.5$, and $3.0$, the plausible range of early-phase ejecta PSDs (Figure \ref{fig:size_distribution}). At this stage, the comparison focused on the relative surface–shadow–plume contrast rather than on an optimized match to the absolute radiance. The case $k = 2.0$ favors the largest particles, producing a plume that is comparatively dark and insufficiently contrasted against the Dimorphos surface. The intermediate case $k = 2.5$ yields an improved brightness profile consistent with estimates for the nascent plume but still fails to reproduce the observed contrast between the ejecta and the surface. The steeper slope, $k=3.0$, most clearly reproduced this relative contrast, although the fixed-density trial overestimated the observed absolute radiance. It was therefore retained for the subsequent adjustment of the number-density profile. The improved fit reflects the need for a non-negligible population of small particles: fine grains increase scattering efficiency near the surface, producing the observed bright ejecta regions while also generating the localized shadowing that forms the dark band adjacent to Dimorphos.

Although steeper than the nominal early-time PSD suggested by \citeA{lolachi2025ejecta}, a slope of $k = 3.0$ remains consistent with the wider set of post-impact PSD constraints, particularly those inferred from the evolution of the dust tail. The preference for a steeper PSD in our near-surface modeling may arise from enhanced radiative influence of fine particles in the optically thick region, early gravitational fallback that selectively removes larger grains near the surface, or unresolved spatial inhomogeneities in the ejecta density. Accordingly, the PSD slope $k = 3.0$ represents a physically motivated and observationally justified choice for reproducing the near-surface brightness structure of the DART ejecta plume in our radiative-transfer framework.

In addition to testing different PSD slopes, we evaluated the influence of the discrete grain-size grid used to approximate the continuous particle size distribution. Our initial calculations included particle radii of 0.5~$\mu$m, 5~$\mu$m, 125~$\mu$m, and 5~mm, spanning sub-micron to millimeter scales. This configuration consistently produced ejecta brightness levels that were systematically higher than observed, resulting in profiles that were overly bright and difficult to reconcile with the LUKE measurements, even when the overall number density was reduced. The excessive radiance indicates that the inclusion of abundant sub-micron particles significantly enhanced the scattering efficiency and optical depth of the near-surface plume. To extend the size range toward the larger fragments indicated by post-impact studies, we incorporated 5~cm grains into the size grid. Although the inclusion of centimeter-scale particles improved the overall radiometric behavior by redistributing scattering contributions toward larger grains, retaining the smallest (0.5~$\mu$m) particles continued to produce excessive near-surface brightness. We therefore adopted a revised size set consisting of 5~$\mu$m, 125~$\mu$m, 5~mm, and 5~cm particles. When combined with a power-law slope of $k = 3.0$, this configuration provided the closest agreement with the LUKE brightness profiles. The final selection reflects a balance between representing the contribution of small-to-intermediate grains and avoiding unrealistically high optical depths that would overpredict plume brightness in the radiative transfer simulations.

\subsection{Number density adjustment} 
\label{SubSec:Number_density}
\begin{figure*}[ht!]
\centering
    \includegraphics[width=\textwidth]{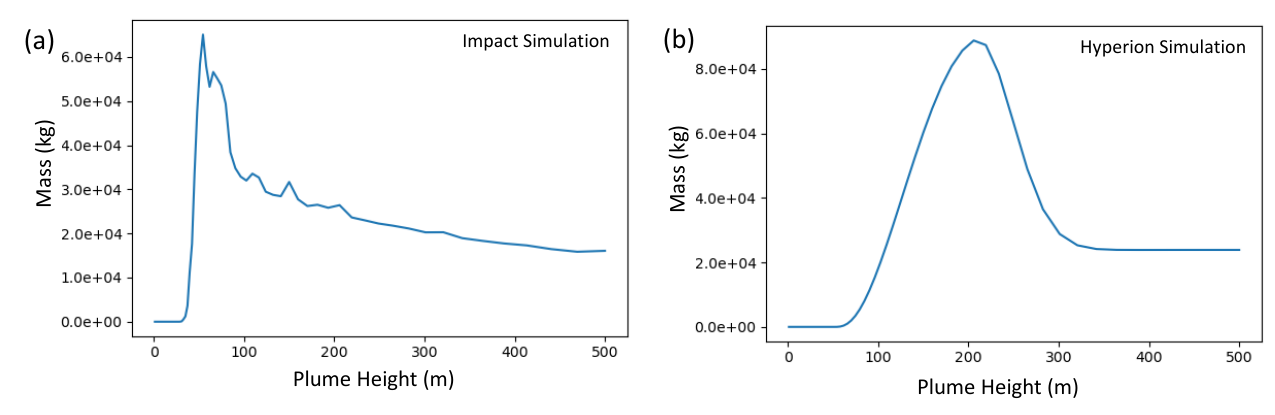}
    \caption{Mass distribution profiles as a function of plume height: (a) the mass distribution inferred from impact simulations by \citeA{raducan2024physical} and (b) the best-fit mass distribution derived from the Hyperion radiative transfer modeling.}
    \label{fig:left}
\label{fig:massdistribution}
\end{figure*}

In this section, we compare radiative transfer simulations based on two number density prescriptions: one derived from the ejecta mass distribution predicted by the impact simulations of \citeA{raducan2024physical}, and another obtained by adjusting this reference profile to reproduce the observed brightness structure in the LICIACube LUKE image at 175~s. Figure~\ref{fig:massdistribution}(a) shows the mass distribution inferred from the impact simulations, while Figure~\ref{fig:massdistribution}(b) shows the distribution corresponding to the best-fit image from our radiative transfer modeling. Following this comparison framework, we adjust the vertical number density distribution. In the Hyperion radiative transfer framework, the number density specifies the local particle concentration and directly controls the optical depth and scattering along each line of sight, thereby governing the modeled brightness distribution of the ejecta plume for a given particle size distribution and scattering properties. In our simulations, we assume that the number density depends only on the distance from the surface of Dimorphos along the plume axis (hereafter referred to as plume height) and does not vary azimuthally.

As the reference model, we adopt the ejecta mass distribution at 175~s after the DART impact derived from the impact simulations of \citeA{raducan2024physical} (Figure~\ref{fig:massdistribution}(a)), which provides the vertical distribution of ejecta mass as a function of height above the surface. We adjust only the overall normalization and the vertical location of the mass peak to reproduce the observed brightness profile in the LICIACube image. For each adjusted mass distribution, we compute the corresponding number density distribution using the assumed bulk density and particle size distribution. The resulting number density profile is used as input to the Hyperion radiative transfer simulations, and the synthetic brightness profiles are compared directly with the LUKE observations, thereby constraining the distribution that best reproduces the observed ejecta brightness structure.

\begin{figure*}[ht!]
\centering
    \includegraphics[width=\textwidth]{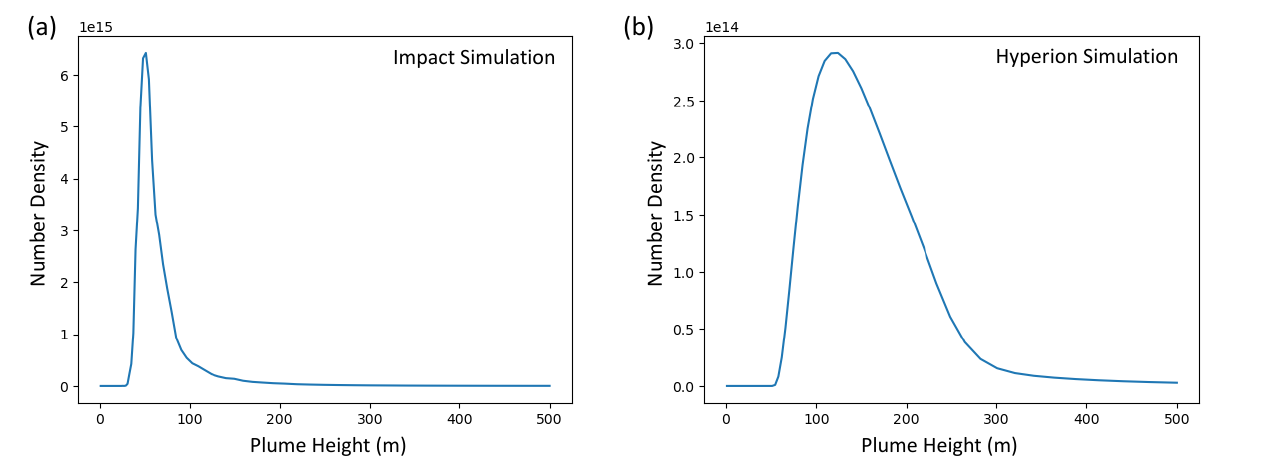}
    \caption{Number density distributions converted from the mass distribution constrained by impact simulations of \citeA{raducan2024physical} (a) and from the mass distribution that produces the best-fit Hyperion-simulated image (b).}
    \label{fig:left}
\label{fig:numberdensity_distribution}
\end{figure*}

\begin{figure*}[ht!]
\centering
    \includegraphics[width=0.8\textwidth]{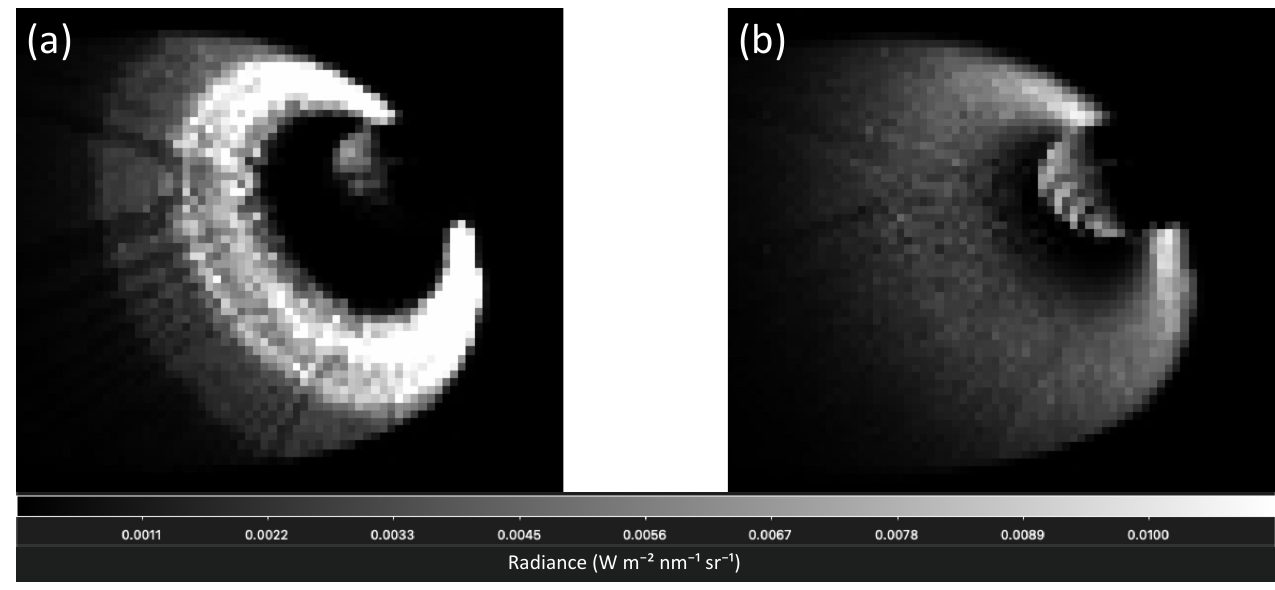}
    \caption{Hyperion-simulated images corresponding to different mass distribution profiles: (a) the mass distribution constrained by impact simulations of \citeA{raducan2024physical} and (b) the best-fit mass distribution derived from the radiative transfer modeling.}
\label{fig:images_mass_distribution}
\end{figure*}

The corresponding number density distributions converted from these mass distributions are shown in Figure~\ref{fig:numberdensity_distribution}(a) and (b), respectively. Although the mass and number density profiles exhibit similar overall trends, they do not have identical shapes. This difference arises because the conversion from mass to number density explicitly accounts for the volume of each plume layer defined by the conical ejecta geometry. The mass distribution $M(h)$ is converted to particle number as $N(h) = M(h)/m_{\mathrm{p}}$, where $m_{\mathrm{p}}$ is the representative particle mass derived from the assumed size distribution and grain density. The number density is then given by $n(h) = N(h)/V_{\mathrm{shell}}(h)$, where $V_{\mathrm{shell}}$ is the volume of a hollow conical shell between successive heights. The rapid increase of this volume with height leads to a more gradual variation of number density compared to the mass distribution. These number density profiles are used as input parameters for the Hyperion simulations, and the resulting synthetic images are presented in Figure~\ref{fig:images_mass_distribution}(a) and (b). Figure~\ref{fig:images_mass_distribution} represents a different stage of the analysis from Figures~\ref{fig:size_distribution}. Figure~\ref{fig:size_distribution} examines sensitivity to the PSD slope using a fixed number-density prescription, whereas Figure~\ref{fig:images_mass_distribution} shows the subsequent adjustment of the vertical number-density profile. This adjustment reduces the excessive radiance of the initial trials and improves the overall surface--shadow--plume contrast. The primary parameters adjusted to reproduce the observed brightness pattern in the LICIACube image are the location of the mass peak along the plume height and its magnitude. Variations in these parameters directly control the spatial extent and position of optically thick regions within the plume. As shown in Figure~\ref{fig:numberdensity_distribution}(a), when the optically thick portion of the plume, corresponding to the peak in number density, is located very close to the surface of Dimorphos, it substantially attenuates incident sunlight before it reaches the surface. This attenuation reduces the amount of solar radiation available for surface scattering. In addition, the dense overlying ejecta enhances geometric shadowing and suppresses multiple scattering between the surface and the plume. Together, these effects lead to a markedly dimmer appearance of Dimorphos in the simulated images, as observed in Figure~\ref{fig:images_mass_distribution}(a). The peak magnitude of the number density also strongly affects the brightness of the ejecta plume. In the impact-simulation–derived number density profile, the peak number density exceeds 6 $\times$ 10$^{15}$ m$^{-3}$, as shown in Figure~\ref{fig:numberdensity_distribution}(a). This produces an ejecta plume that is excessively bright and inconsistent with the LICIACube observations. By adjusting both the height at which the mass peak occurs and its magnitude, as shown in Figure~\ref{fig:numberdensity_distribution}(b), we obtain a number density profile with a lower peak value of $3 \times 10^{14},\mathrm{m^{-3}}$ and a more gradual increase and decrease with plume height compared to the impact-simulation–derived profile shown in Figure~\ref{fig:numberdensity_distribution}(a). In contrast to the sharply peaked profile in Figure~\ref{fig:numberdensity_distribution}(a), which exhibits a rapid rise and fall in number density over a narrow height range, the adjusted profile produces a broader vertical extent of moderate optical depth within the plume.

Additionally, the location of the optically thickest region plays a critical role in shaping shadowing effects, including the dark region between the ejecta plume and the surface of Dimorphos as well as shadowed areas on the surface itself. In the impact-simulation–derived profile, the peak number density occurs at a plume height of approximately 80~m, placing the optically thickest portion of the plume very close to the surface of Dimorphos. In our adjusted profile, the peak is shifted upward to a plume height of approximately 120~m. This upward displacement reduces excessive attenuation and shadowing near the surface while maintaining a sufficiently dense region within the plume to reproduce the observed ejecta brightness. Locating the optically thickest region slightly farther from the surface therefore helps preserve a reasonable contrast between the ejecta and the illuminated portions of Dimorphos, avoiding a surface brightness pattern that is inconsistent with the LICIACube observations.
As a result, the impact-simulation–derived number density profile, with peak values exceeding $6 \times 10^{15},\mathrm{m^{-3}}$, produces an ejecta plume that is excessively optically thick and significantly brighter than observed in the LICIACube data as seen in Figure~\ref{fig:images_mass_distribution}(a). In contrast, the adjusted number density profile yields synthetic images that more closely reproduce the observed brightness structure of the ejecta plume, as shown in Figure~\ref{fig:images_mass_distribution}(b).

\subsection{Brightness Profile Comparison and Best-fit Image}
\label{SubSec:BrightnessComparison}
\begin{figure*}[h!]
\centering
    \includegraphics[width=\textwidth]{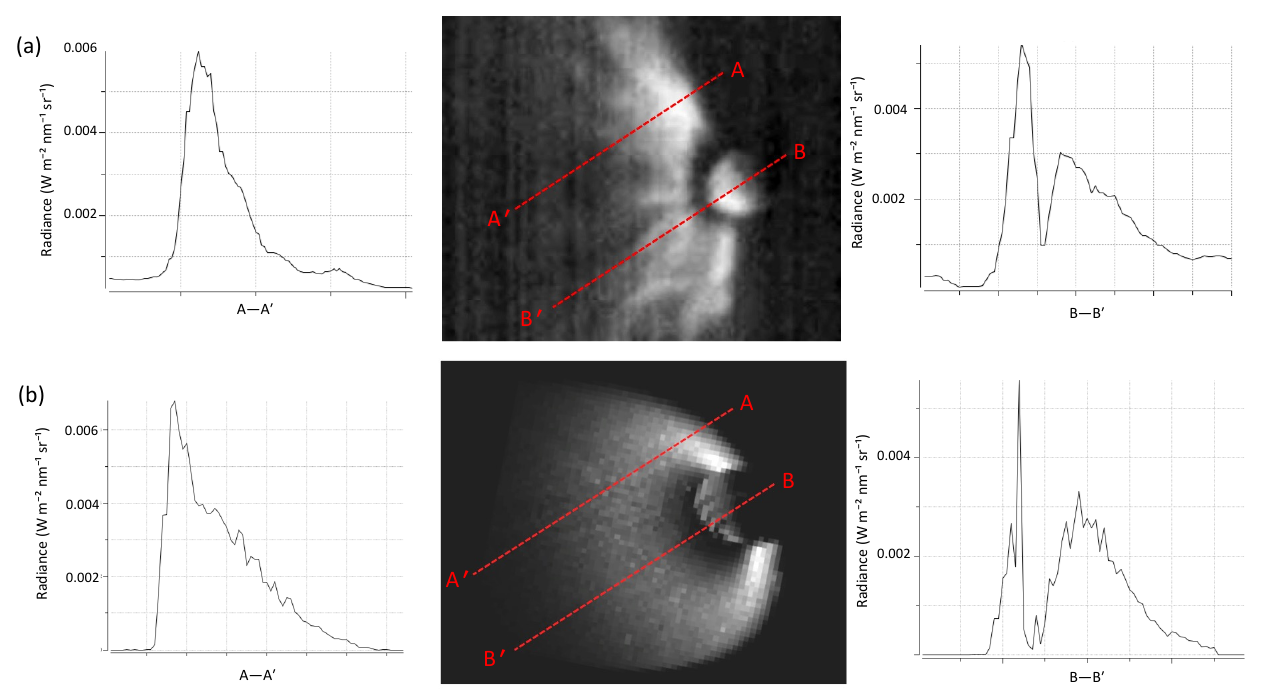}
\caption{Comparison of brightness profiles between (a) the LICIACube observations and (b) the Hyperion-simulated image. Brightness profiles, expressed as spectral radiance ($W m^{-2} nm^{-1} sr^{-1}$), were extracted from images displayed with linear intensity scaling in SAOImage DS9. The A–A' cut traces the brightest region along the edge of the ejecta cone. The B–B' cut passes through the center of Dimorphos and the ejecta cone, intersecting the bright surface of Dimorphos and the darker band between the surface and the ejecta cone.}
\label{fig:HyperionImage}
\end{figure*}

To evaluate the combined effects of the adopted PSD slope ($k = 3.0$) and the adjusted number density profile, we directly compare brightness profiles extracted from the LUKE red-filter image (Figure~\ref{fig:HyperionImage}(a)) with those derived from the corresponding best-fit Hyperion simulation (Figure~\ref{fig:HyperionImage}(b)). Two representative cuts were selected for profile extraction, indicated by red-dotted lines in the images. The A–A' cut traces the region of maximum brightness within the ejecta plume, while the B–B' cut extends from the center of Dimorphos into the ejecta cone, intersecting the illuminated surface, the intervening dark band, and the bright plume. Figure~\ref{fig:HyperionImage} presents the extracted brightness profiles for both the measured (top panel) and modeled (bottom panel) ejecta, expressed in spectral radiance units ($W m^{-2} nm^{-1} sr^{-1}$).

First, along the A–A' cut, both the LUKE observation and the simulation exhibit a rapid rise in brightness to a peak value of approximately $0.006$–$0.007~W\,m^{-2}\,nm^{-1}\,sr^{-1}$, followed by a gradual decline with increasing distance from the surface. The agreement in peak radiance and overall amplitude indicates that the adopted particle size distribution and vertical number density structure reproduce the optical depth of the near-surface plume with reasonable accuracy. A minor discrepancy appears in the post-peak behavior: the simulated profile decreases more smoothly than the observed one. This difference likely reflects the assumption of azimuthal symmetry and a vertically stratified but laterally uniform number density. In contrast, the actual ejecta plume likely contains small-scale inhomogeneities and localized density variations that enhance brightness fluctuations in the LUKE data.

Along the B–B' cut, the profile captures the relative radiometric contrast between the illuminated surface of Dimorphos, the adjacent dark band, and the ejecta cone. The simulation successfully reproduces the high surface radiance, the pronounced attenuation within the dark region, and the subsequent increase in brightness toward the plume. This behavior confirms that the adopted vertical density distribution produces an optically thickest region capable of generating both attenuation and geometric shadowing consistent with the observations. Remaining differences are primarily confined to the sharpness and depth of the dark band. In the LUKE image, the transition from the bright region to the shadowed region is more abrupt, whereas the simulated profile exhibits a smoother gradient. This likely arises from simplified geometric assumptions in the model, including axisymmetry and the absence of three-dimensional plume irregularities, which can amplify localized shadowing and contrast in the observed image. With this best-fit configuration, the total ejecta mass contained within the plume up to a height of 500~m is estimated to be $1.25 \times 10^{6}$~kg, calculated by integrating the number density distribution over the plume volume and converting particle number to mass using the adopted particle size distribution and bulk density.

\section{Discussion and Conclusions}
\label{Sec:Discussion_Conclusions}
By applying three-dimensional radiative transfer modeling to the LUKE image acquired 175~s after impact, we were able to directly probe the structure of the near-surface DART ejecta plume. Through iterative adjustment of the particle size distribution and vertical number density profile, we identified the best-fit configuration consisting of a power-law slope $k = 3.0$, a density peak at approximately 120~m above the surface, and a total ejecta mass of $1.25 \times 10^{6}$~kg contained within 500~m of plume height. With this configuration, we successfully reproduce the key large-scale brightness features observed in the LUKE image, including the approximate peak radiance of the plume and the surface–shadow–plume contrast formed by the illuminated surface of Dimorphos, the adjacent dark band, and the bright ejecta. Although some localized differences remain in the detailed appearance of Dimorphos and the center of the ejecta cone, the model captures the principal radiometric structure of the near-surface plume.

When we compare our mass estimate with previous studies (summarized in Table~\ref{tab:mass_estimate_summary}), an apparent difference emerges. Numerical impact simulations using the Bern SPH shock-physics code \cite{raducan2024physical} yield total ejecta masses of $(1.3$--$2.2) \times 10^{7}$~kg, and point-source scaling relations calibrated to these simulations give $(2.34$--$4.06) \times 10^{7}$~kg \cite{cheng2023momentum}. Photometric analyses of LICIACube images under optically thin assumptions produce lower-limit values of $(0.85$--$1.19) \times 10^{7}$~kg \cite{lolachi2025ejecta}. Accounting for the contribution from optically thick ejecta, \citeA{lolachi2025ejecta} derive a revised total mass estimate of $(1.6 \pm 0.3) \times 10^7$~kg using an empirical correction. Our value is roughly an order of magnitude smaller; however, this difference is largely a consequence of scope. We restrict our analysis to ejecta within 500~m of the surface, whereas previous studies aim to estimate the total escaping mass. The region we model corresponds to the optically thick region of the plume at 175 s after impact, during the active excavation stage prior to the full development of the expanding ejecta cloud.

\begin{table}
\caption{Summary of published ejecta mass estimates derived from observations 175 s after impact}
\label{tab:mass_estimate_summary}
\centering
\begin{tabular}{l p{0.48\textwidth} l}
\hline
Total Mass (kg) & Description & Reference \\
\hline
(1.3--2.2)$\times10^{7}$ 
& Numerical impact simulations using the Bern SPH shock physics code. 
& \citeA{raducan2024physical} \\

(2.34--4.06)$\times10^{7}$ 
& Point source scaling models fitted to numerical simulations of the DART impact into a rubble pile Dimorphos. 
& \citeA{cheng2023momentum} \\

$(1.6 \pm 0.3) \times 10^{7}$ & Photometric analysis of LICIACube plume images; lower-limit estimate of $(0.85$--$1.19) \times 10^{7}$~kg from optically thin assumptions, with an empirical correction applied to account for optically thick ejecta.
& \citeA{lolachi2025ejecta} \\

$\sim1.25\times10^{6}$\\(within 500 m)
& Three dimensional radiative transfer modeling of LUKE images constraining the particle size distribution and vertical number density within 500 m of the surface. 
& This study \\
\hline
\end{tabular}
\end{table}

\subsection{Differences Between Radiative and Dynamical Approaches}

To enable a more direct comparison with the impact simulations of \citeA{raducan2024physical}, we evaluated the SPH-derived mass distribution within the same restricted height range considered in our radiative transfer modeling. By integrating their mass profile up to a plume height of 500~m (Figure~\ref{fig:massdistribution}(a)), we obtain a near-surface ejecta mass of $1.29 \times 10^{6}$~kg. Our best-fit value of $1.25 \times 10^{6}$~kg is therefore slightly lower but fully consistent within this limited region.

Although the total mass within 500~m is comparable, the vertical mass distributions inferred from the two approaches exhibit noticeable differences, as seen in Figure~\ref{fig:massdistribution}. In the SPH simulation, the mass profile rises more sharply near the surface and peaks at lower altitude, while our radiative transfer–derived distribution shows a broader peak shifted upward. This difference reflects the distinct physical sensitivities of the two methods. Our radiative transfer modeling is intrinsically weighted toward particles that dominate the scattering cross section—primarily micron- to millimeter-scale grains that control the optical depth and shadowing behavior of the plume. Impact simulations, in contrast, track the full fragment population and are especially sensitive to larger fragments that contribute substantially to the total mass but only modestly to optical brightness. Because brightness scales with scattering cross section rather than mass, our method favors distributions that reproduce the observed radiance structure, even if the underlying mass is redistributed vertically relative to the dynamical solution.

Additional discrepancies likely arise from geometric and material assumptions. In our framework, the opening angle and vertical density profile of the ejecta plume directly determine the mapping between mass and optical depth. A peak that is too close to the surface of Dimorphos produces excessive attenuation and unrealistically dim surface brightness, requiring a slight upward shift of the optically thickest region to match the LUKE observations. Impact simulations, meanwhile, depend sensitively on assumed cohesion, internal friction, and porosity, which influence the excavation efficiency and near-surface mass concentration.

Taken together, these differences highlight the complementary nature of radiative and dynamical modeling. While impact simulations constrain the total excavated mass and momentum transfer, radiative transfer modeling constrains the optical structure and particle size distribution of the lower part of the ejecta. The close agreement in integrated mass within 500~m strengthens confidence that both approaches are capturing consistent aspects of the DART impact, even if some details of the vertical distributions differ.

\section*{Data Availability Statement}

The Dimorphos shape model (version~3.0) is available from the NASA Planetary Data System (PDS) as described by \citeA{daly2023derived} at \url{https://doi.org/10.26007/er6h-qf14}. Geometric and spacecraft trajectory information were obtained from the DART SPICE kernels; the metakernel used was \textit{dart\_tlf\_v01.tm}. The kernel archive is available from the PDS and NAIF as described by \citeA{nair2025dartspice} at \url{https://doi.org/10.17189/9acz-aw83}. The calibrated LICIACube LUKE image is available from the PDS Small Bodies Node as described by \citeA{farnham2025luke} at \url{https://doi.org/10.26007/5vhp-pe80}. The radiative transfer simulations used Hyperion, version~\textit{v0.9.11}, as described by \citeA{robitaille2011hyperion}. The software is available at \url{https://github.com/hyperion-rt/hyperion} and cited as \citeA{hyperionSoftware}.

\acknowledgments

Y.K. and L.K. acknowledge that this work was supported by the NASA DART PSP grant $\#$80NSSC21K1131. This work was also supported by the DART mission, NASA Contract No. 80MSFC20D0004. R.L. and D.A.G. acknowledge work done through the Center for Research and Exploration in Space Science and Technology (CRESST-II) supported by NASA award No. 80GSFC24M0006. R.L., D.A.G., and T.J.S. were supported by NASA/GSFC Internal Scientist Funding Model (ISFM) Exospheres, Ionospheres, Magnetospheres Modeling (EIMM). T.J.S. was also supported by NASA Solar System Exploration Research Virtual Institute (SSERVI) Lunar Environment and Dynamics for Exploration Research (LEADER).

\section*{Author Contributions}

Y.K. and L.K. carried out all primary aspects of the study, including conceptualization, modeling, analysis, interpretation, and manuscript preparation. T.L.F. provided the calibrated LICIACube images and geometric information used in the simulations. R.L., D.A.G., and T.J.S. provided the light-scattering properties. S.D.R. provided the ejecta mass distribution profiles from impact simulations used to initialize the modeling. All authors contributed to reviewing and editing the manuscript.

\section*{Conflict of Interest Statement}
The authors have no conflicts of interest to disclose.

\bibliography{Reference}

%Reference citation instructions and examples:
%
% Please use ONLY \cite and \citeA for reference citations.
% \cite for parenthetical references
% ...as shown in recent studies (Simpson et al., 2019)
% \citeA for in-text citations
% ...Simpson et al. (2019) have shown...
%
%
%...as shown by \citeA{jskilby}.
%...as shown by \citeA{lewin76}, \citeA{carson86}, \citeA{bartoldy02}, and \citeA{rinaldi03}.
%...has been shown \cite{jskilbye}.
%...has been shown \cite{lewin76,carson86,bartoldy02,rinaldi03}.
%... \cite <i.e.>[]{lewin76,carson86,bartoldy02,rinaldi03}.
%...has been shown by \cite <e.g.,>[and others]{lewin76}.
%
% apacite uses < > for prenotes and [ ] for postnotes
% DO NOT use other cite commands (e.g., \citeA, \cite, \citeyear, \nocite, \citealp, etc.).
%

\end{document}